\documentclass{article}
\usepackage{epsf}
\usepackage{lipsum}
\usepackage{caption}
\usepackage{amssymb,ComplexSystems}
\usepackage{graphicx}
\usepackage{epstopdf}
\usepackage{amsfonts}
\usepackage{url}
\usepackage{amsmath}

\begin{document}
\title{ New Directions In Cellular Automata}
\author{\authname{Abdulrhman Elnekiti}\\[2pt] 
\authadd{Department of Computer Science} \\{ University of Turkish  Aeronautical  Association}\\
\authadd{11 Bahcekapi, 06790 Etimesgut - Ankara, Turkey} \\
\authadd{Email : Abdulrhman.Elnekiti@stu.thk.edu.tr} } 
\maketitle
\begin{abstract}
We Propose A Novel Automaton Model which uses Arithmetic Operations as the Evolving Rules, each cell has the states of the Natural Numbers $k$ = $(\mathbb{N})$, a radius of  $r$ = $\frac{1}{2}$ and operates on an arbitrary input size. The Automaton reads an Arithmetic Expression as an input and outputs another Arithmetic Expression. In Addition, we simulate a variety of One Dimensional Cellular Automata Structures with different Dynamics including Elementary Cellular Automata. \end{abstract}
\section{History  \& Overview}
A Cellular Automaton (CA) is a Discrete Dynamical System which consists of Identical Cells, each cell has a defined number of states and the cells evolve according to a local rule in a defined number of iterations. CA was introduced by Von Neumann and Ulam \cite{text-a} to model natural growth processes like Seashell and Snowflakes \cite{text-b}. \\\\ Stephen Wolfram has explored One-Dimensional CA where each cell has the states of [1 or 0], known as Elementary Cellular Automata (ECA) \cite{text-c}. We propose an Automaton model that shows a variety of structures based on a specific arithmetic expression formed as the input.
\section{Formal Description}
A One-Dimensional Cellular Automaton consists of a finite row of cells, we define a cell $ c $ and the Right-Most Neighboring cell as $c + 1$, a radius $r$ and the possible states for each cell as $k$. The Automaton can be described by $r$ = $\frac{1}{2}$ and $k$ = $\mathbb{N}$, where it depends on the current cell $c$ and the next cell to it $c+1$, and each cell has the states of the Natural Numbers. \\\\
\textbf{Transition Function} : Let $c(i,t)$ denote the state of the $i$th cell at time $t$, the cell state in the next time step is defined by $c(i,t+1)$ using the transition function $\delta$ :
\begin{align} 
c(i,t+1) = \delta [c (i,t), c (i+1,t)] \\
\delta = |  c(i,t) - c(i+1,t)  | \\
c (i,t+1) = | c(i,t) - c(i+1,t) |
\end{align}\\ 
\textbf{Input Formation.}  Let $P$ denote the input, $P \subset \lbrace \mathbb{N}, -\rbrace (*)$ where $\mathbb{N}$ is the set of The Natural Numbers, $(-)$ is the Subtraction Operation and the Asterisk defines a repetition of the Sequence of the elements, i.e. $P$ = ${\left[ 2-0-1-4\right] }$, $P$ = ${\left[2-0-1-5\right] }$ or $P$ = ${\left[2-0-1-6\right] }$, for a better interpretation we define $P$ by default as $P$ = ${\left[2-0-1-7-0-4-7-8-9-0-9-8-7-4-0-7-1-0-2\right].}$ \\\\
\centerline{\includegraphics[scale=0.25]{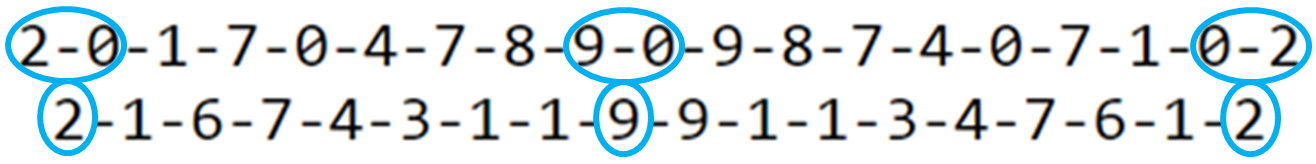}}
\captionof{figure}{The Input $P$ And the First Resulting Iteration Written Under It Using the transition function $\delta$\\}
The blue circles describe that the resulting cell of any Two Continuous Cells is placed under it. \\ \\ \centerline{2-0-1-7-0-4-7-8-9-0-9-8-7-4-0-7-1-0-2}\\
\centerline{2-1-6-7-4-3-1-1-9-9-1-1-3-4-7-6-1-2}\\
\centerline{1-5-1-3-1-2-0-8-0-8-0-2-1-3-1-5-1}\\
\centerline{4-4-2-2-1-2-8-8-8-8-2-1-2-2-4-4}\\
\centerline{0-2-0-1-1-6-0-0-0-6-1-1-0-2-0}\\
\centerline{2-2-1-0-5-6-0-0-6-5-0-1-2-2}\\
\centerline{0-1-1-5-1-6-0-6-1-5-1-1-0}\\
\centerline{1-0-4-4-5-6-6-5-4-4-0-1}\\
\centerline{1-4-0-1-1-0-1-1-0-4-1}\\
\centerline{3-4-1-0-1-1-0-1-4-3}\\
\centerline{1-3-1-1-0-1-1-3-1}\\
\centerline{2-2-0-1-1-0-2-2}\\
\centerline{0-2-1-0-1-2-0}\\
\centerline{2-1-1-1-1-2}\\
\centerline{1-0-0-0-1}\\
\centerline{1-0-0-1}\\
\centerline{1-0-1}\\
\centerline{1-1}\\
\centerline{0}
\captionof{figure}{The Complete Iterations Generated From The Input $P$ Using The Transition Function $\delta$\\}
To show a complex structure in the automaton, we define a pattern $S$ that will be highlighted in the iterations. The pattern is defined as a sequence of the Natural Numbers  $\mathbb{N}$ and the Subtraction Operation $S \subset \lbrace \mathbb{N}, -\rbrace\left( *\right) $, i.e. $S$ = $\left[ 0-\right]$, $S$ = $\left[ 1-0\right]$, $S = \left[ 1-0-1\right] \ldots  etc$ \\\\
\centerline{\includegraphics[scale=0.33]{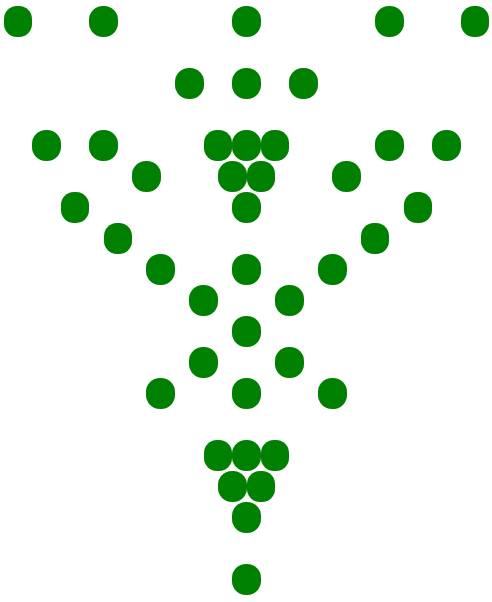}}
\captionof{figure}{A Complex Structure Generated Using The Default Input $P$ And $S$ = $\left[0-\right]$ \\}
\section{Simulation of Elementary Cellular Automata}
In this section we will simulate different structures from ECA using the Proposed Automaton. For each input of the following simulations, it can be found in the appendices and will be referred to as $Simulation$ $-$ $1 = (A.1) \ldots Simulation$ $-$ $2 = (A.2) \ldots etc$. \\\\\\
\centerline{\includegraphics[scale=0.69]{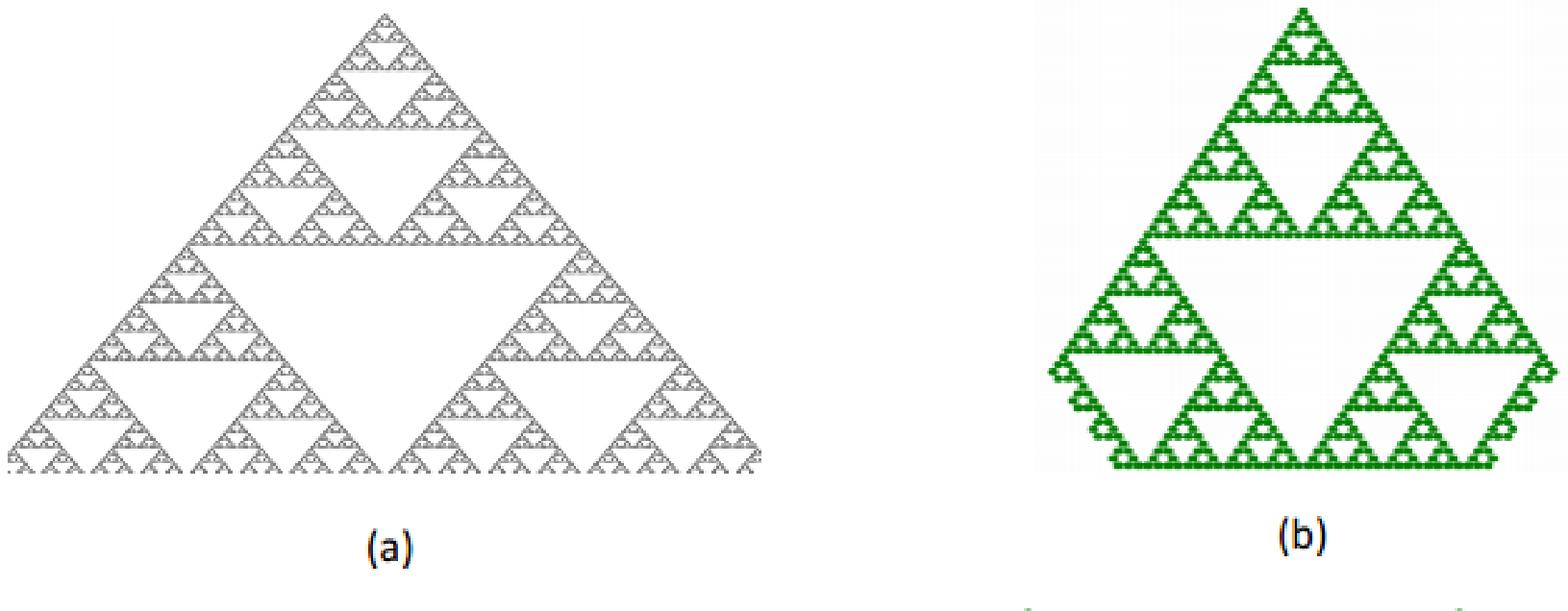}}
\centerline{\includegraphics[scale=0.66]{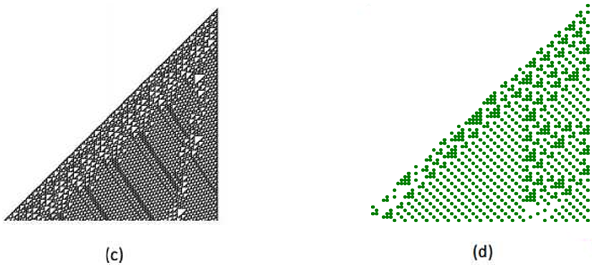}}
\centerline{\includegraphics[scale=0.97]{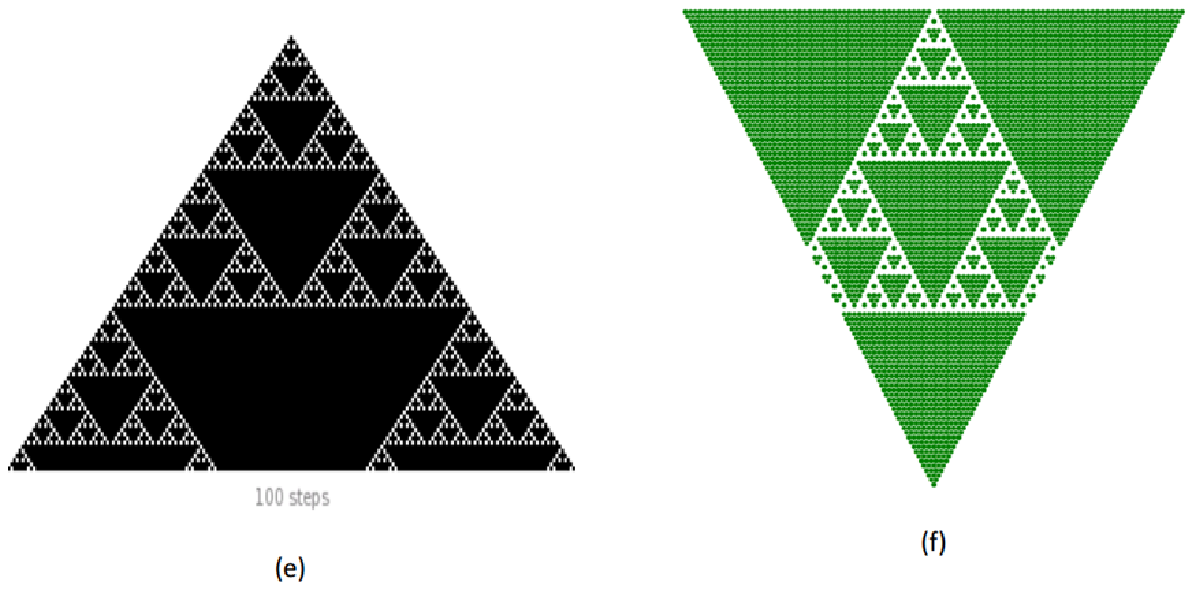}}
\caption{A Comparison Between ECA Structures And Their Equivalent Simulations \\}
The figure shows a comparison between different structures of ECA on the left side as $(a), (c)$  $\&$ $(e)$, known also as Rule$-$90 \cite{website1} , Rule$-$110 \cite{website2} $\&$ Rule$-$182 \cite{website3} respectively. Where on the right side $(b), (d)$ $\&$ $(f)$ shows the corresponding simulations using the proposed automaton. \\\\
\textbf{Properties Of The Automaton. }The Automaton can simulate multiple structures from ECA with the same input applied, i.e. Rule$-$90 shown in \textbf{Figure 4. (a)}, it can be simulated using $P = (A.1)$ and $S =(1-)$ highlighted. For Rule$-$182 shown in \textbf{Figure 4. (e)}, it can be simulated using the same input $P = (A.1)$ and $S =\left(0-\right)$ highlighted.\\\\
Another property is simulating Rule$-$110 which has been proven to be Turing-Complete \cite{a-review} shown in \textbf{Figure 4. (c)}, the corresponding simulation is \textbf{Figure 4. (d)} using $P = (A.2)$ and $S = (1-)$ .  \\\
\textbf{Symmetric Vs Non-Symmetric Structures.} An interesting property of the automaton is transforming a Non-Symmetric structure to a Symmetric one based on the input modification. \\\\This is done by combining the original input $P$  (where $P = P_{old})$ with the reverse of it described as $P_{new} = (P_{old})(Reverse(P_{old}))$, i.e. $P_{old} = \left[1-5\right]$ $\&$ $Reverse(P_{old})= \left[5-1\right]$, which will give us $P_{new}$ = $\left[1-5-5-1\right]$. \\\\
\centerline{\includegraphics[scale=0.49]{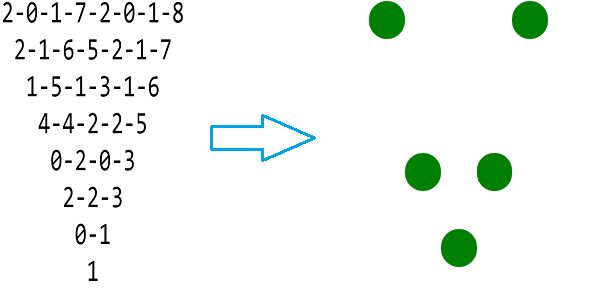}}
\caption{A Structure Generated Using $P_{1} = (2-0-1-7-2-0-1-8)$ And $S = [0-]$\\}
$\textbf{Figure 5.}$ is symmetric when $P_{1}$ is modified to include itself and the reverse of it as $P_{1 new} = (2-0-1-7-2-0-1-8) - (8-1-0-2-7-1-0-2)$, it is equivalent to $P_{1 new} = (2-0-1-7-2-0-1-8-8-1-0-2-7-1-0-2)$. \\\\\\
\centerline{\includegraphics[scale=0.29]{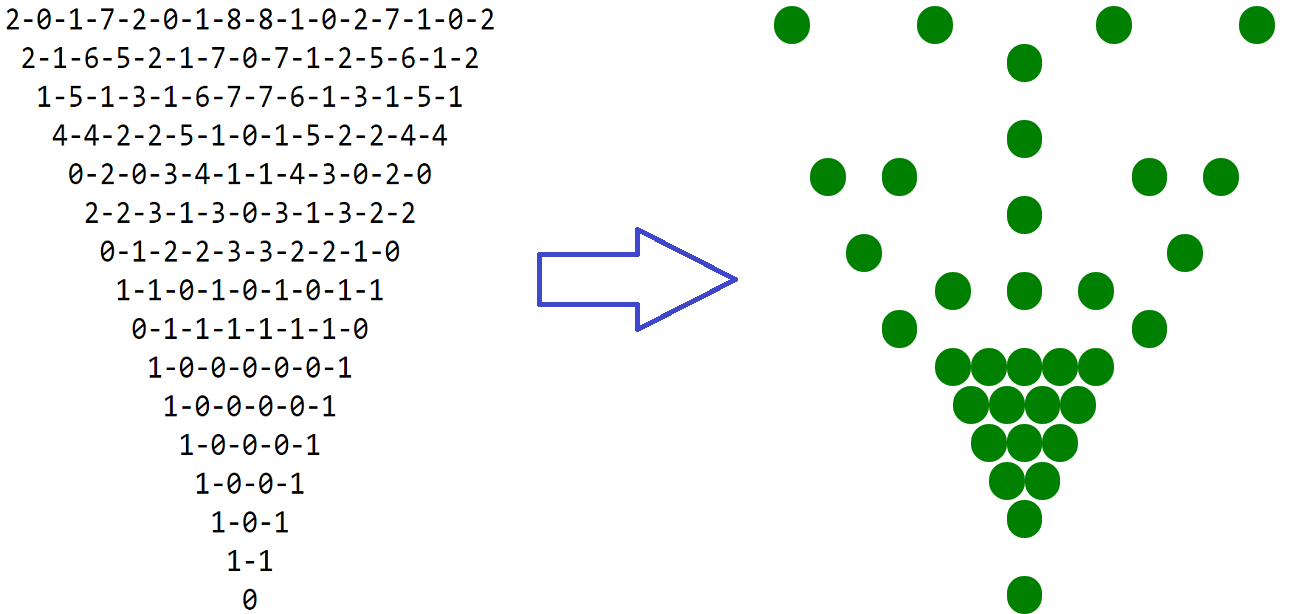}}\\
\caption{The Symmetric Structure Based On $P_{1 new}$ and $S = \left[0-\right]$} \vspace{\baselineskip}\vspace{\baselineskip} While highlighting $S$ = $\left[0-\right]$ shows a complex structure, other patterns can show different structures as well, i.e. $S = \left[1-\right]$ or $S = \left[2-\right]$.\\\\\\ 
\centerline{\includegraphics[scale=0.22]{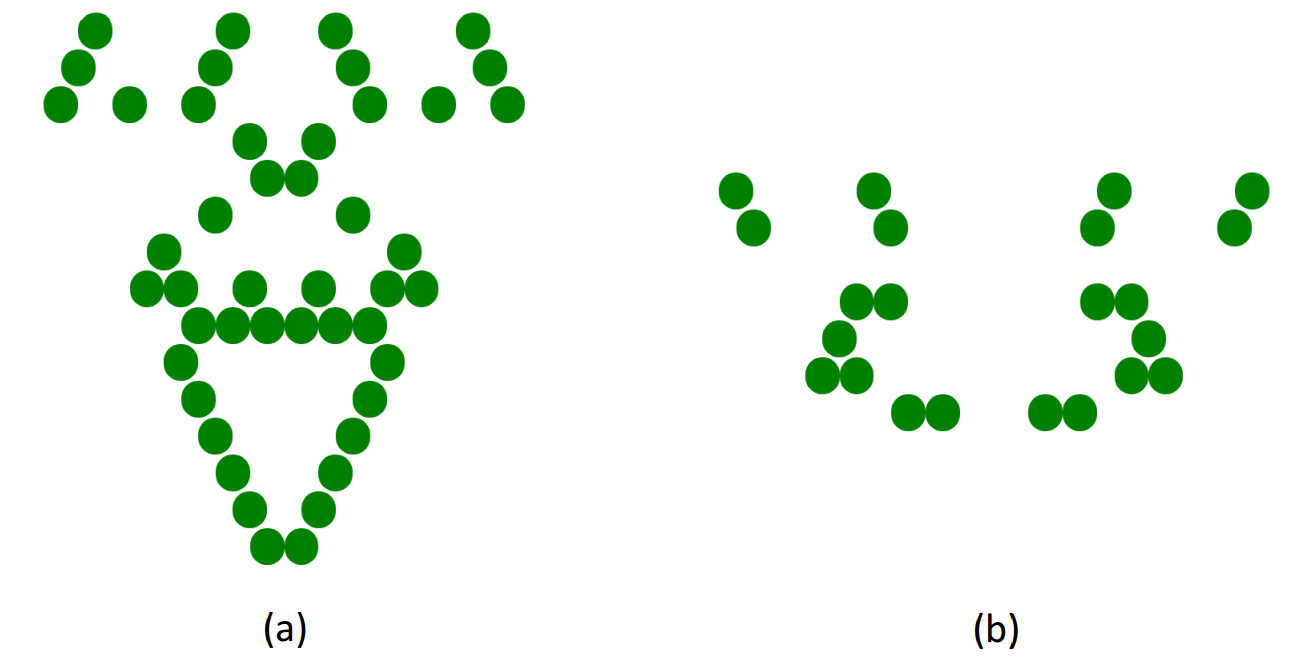}}
\caption{Different Structures Based On $P_{1 new}$ with different $S$ highlighted in $(a)$ $\&$ $(b)$ \\} \textbf{Figure} \textbf{7.} shows different structures as $(a)$ $\&$ $(b)$ generated from $P_{1 new}$ where $(a)$ has $S=\left[1-\right]$ and $(b)$ has $S=\left[2-\right]$.
\section{Conclusion \& Future Work} 
The purpose of this paper was to propose an automaton model that uses arithmetic operations as the evolving rules and using the natural number as the cells. \\\\There are other properties, concepts and results to explain about the automaton which needs separate research papers, Such As:  \\

1) Simulation of 2$-$Dimensional CA \\
2) Classification Of The Structures Generated In It  \\
3) Using Arithmetics $\&$ Algebraic Equations as The Evolving Rules\\
4) Provisional Ideas On How To Integrate it With Boolean Algebra  \\
\appendix 

\begin{enumerate}
\item \centering $0-0-0-0-0-0-0-0-0-0-0-0-0-0-0-0-0-0-0-0-0-0-0-0-0-0-0-0-0-0-0-0-0-0-0-0-0-0-0-0-0-0-0-0-0-0-0-0-0-0-1-0-0-0-0-0-0-0-0-0-0-0-0-0-0-0-0-0-0-0-0-0-0-0-0-0-0-0-0-0-0-0-0-0-0-0-0-0-0-0-0-0-0-0-0-0-0-0-0-0-0-$ \label{haha}\\
\item \centering $9-9-1-0-5-0-9-8-8-9-7-8-9-7-8-9-7-8-9-7-8-9-7-8-9-7-8-9-7-8-9-7-8-9-7-8-9-7-8-9-7-8-9-7-8-9-7-8-9-4-5-1-7-8-9-7-8-9-7-8-9-7-8-9-7-8-9-$ \label{haha2}
\end{enumerate}

 \end{document}